%% file: main.tex
\documentclass{article}
\usepackage[margin = 1.25in]{geometry}

\usepackage{amsthm,amsmath,amsfonts,amssymb}
\usepackage{graphicx}
\usepackage{mathtools}
\usepackage{multirow}
\usepackage{epsfig,url}
\usepackage[utf8]{inputenc} 
\usepackage[ruled]{algorithm2e}

\usepackage{subfiles}

\usepackage{natbib}

\newcommand{\bigCI}{\mathrel{\text{\scalebox{1.07}{$\perp\mkern-10mu\perp$}}}}

\numberwithin{equation}{section}
\theoremstyle{plain}

\theoremstyle{remark}

\newcommand{\paperTitle}{Generalized coarsened confounding for causal effects: a large-sample framework}

\renewenvironment{abstract}%
{%
  \vskip 0.075in%
  \centerline%
  {\large\bf Abstract}%
  \vspace{0.5ex}%
  \begin{quote}%
}
{
  \par%
  \end{quote}%
  \vskip 1ex%
}

\begin{document}

\title{\huge \paperTitle}
\date{}
\author{
\textbf{Debashis Ghosh}\\
\normalsize Department of Biostatistics and Informatics \\
\normalsize Colorado School of Public Health \\
\normalsize \texttt{debashis.ghosh@cuanschutz.edu}
\and 
\textbf{Lei Wang}\\
\normalsize Department of Biostatistics and Informatics \\
\normalsize Colorado School of Public Health \\
\normalsize \texttt{lei.2.wang@cuanschutz.edu}
}

\maketitle

\begin{abstract}
There has been widespread use of causal inference methods for the rigorous analysis of observational studies and to identify policy evaluations.   In this article, we consider a class of generalized coarsened procedures for confounding.  At a high level, these procedures can be viewed as performing a clustering of confounding variables, followed by treatment effect and attendant variance estimation using the confounder strata.   In addition, we propose two new algorithms for generalized coarsened confounding.   While ~\cite{iacus2011multivariate} developed some statistical properties for one special case in our class of procedures, we instead develop a general asymptotic framework.    We provide asymptotic results for the average causal effect estimator as well as providing conditions for consistency.  In addition, we provide an asymptotic justification for the variance formulae in~\cite{iacus2011multivariate}.   A bias correction technique is proposed, and we apply the proposed methodology to data from two well-known observational studies. 
\end{abstract}

\section{Introduction}
\label{sec:introduction}

\subfile{sections/introduction}

\section{Background and Preliminaries}
\label{sec:background}

\subfile{sections/background}

\section{Main Results}
\label{sec:main}
\subfile{sections/methods}


%

\section{Examples}
\label{sec:examples}
\subfile{sections/examples}

\section{Discussion}
\label{sec:discussion}

\subfile{sections/discussion}

\section*{Acknowledgments}
The authors acknowledge funding support from the following sources: the National Institutes of Health, the National Science Foundation, the Veterans Administration, and the Grohne-Stepp Endowment from the University of Colorado Cancer Center.  

\section*{Appendix}

\subfile{sections/appendix}

\bibliographystyle{chicago}
\bibliography{biblio-IB.bib}

\end{document}

%% file: sections/introduction.tex
\noindent 

The use of causal inference methods is expanding across a variety of domains in society.  While their basis lies in fields such as sociology, economics, statistics, biostatistics and computer science, recent applications of causal inference methods have been to topics such as estimating the effect of government policies on COVID19 rates~\citep{hsiang2020effect,chernozhukov2021causal}, evaluation of behavior as a mediator of climate on transmission of the SARS-CoV2-Virus\cite{grover2024does}. 

One of the foundational frameworks for causal inference has been the potential outcomes model \cite{neyman1923applications,rubin1974estimating}.  This framework posits counterfactual outcomes on an individual level used to define causal effects.  In addition, the authors provide a set of assumptions needed to guarantee the identifiability of individual-level causal effects using observed data.    

In the potential outcomes framework, \citet{rosenbaum1983central} were able to show how the propensity score, defined as the probability of treatment given confounders, plays a key role in causal effect estimation and inference with observational data. Under a strong ignorability assumption, the propensity score removes bias attributable to confounding due to its property as a balancing score \citep{rosenbaum1983central}. Defining causal effects using potential outcomes and using the propensity score allows for a two-stage approach to causal effect estimation.  In the first stage, the propensity score is modeled, while at the second stage, the causal effect is estimated in which the propensity score is incorporated.   Examples of propensity score adjustment includes matching, inverse probability weighted estimation, subclassification and hybrid approaches, such as augmented inverse probability weighted estimation; further details about these can be found in \cite{imbens2015causal}.

This article is motivated by an alternative approach to confounder adjustment, termed coarsened exact matching~\citep{iacus2011multivariate}, which is described in \S 2.2.  One of the primary aims of their method was to eliminate the iterative step of re-matching participants until an acceptable amount of balance is achieved. Coarsened exact matching is quite simple in nature and proceeds using the following high-level heuristic:
\begin{enumerate}
\item For each confounding variable, coarsen it into a certain number of categories;
\item Create strata based on the possible combinations of the coarsened values;
\item Compute a causal effect by comparing the outcomes of the treatment groups within the strata.
\end{enumerate}
The theoretical justification provided by \citet{iacus2011multivariate} for coarsened exact matching is a concept they term monotonic imbalance.  They show that bounding the distance between confounders to be small leads to matching procedures that are more flexible than procedures based on the equal percent bias reduction theory developed by Rubin and collaborators \citep{rubin1976multivariate, rubin1992affinely, rubin2006affinely}. One of the main advantages of coarsened exact matching is that it becomes amenable to large-scale database querying approaches to peforming causal inference: see \citet{salimi2016zaliql} as well as \citet{wang2017flame}.

In many practical settings, we will have large numbers of confounding variables, say 50 or more.   For these settings, coarsened exact matching will suffer from some issues.  First, there is a chance that the common support assumption between treatment and control groups will be violated.   Second, the coarsened exact matching algorithm will potentially overcoarsen, leading to strata in which there are no observations, or observations from only one group.  In the latter scenario, the resulting strata are discarded from the calculation of the treatment effect as well as the attendant standard error.   Based on our analysis in Section~\ref{sec:methods}, we find that the coarsened exact matching causal effect estimator is in fact a {\bf data-adaptive strata-based estimator}.  This means that the confounders are used to built strata, from which responses in the treatment and control groups are compared.  This representation also allows us to study the asymptotic properties of the causal effect estimator and its variance more carefully than has been previously done in the literature.

In this article, we develop a new theoretical framework we term generalized coarsened confounding, of which coarsened exact matching is a special case.  Unlike what is presented in~\cite{iacus2011multivariate}, our approach to inference is based on large-sample theory and thus takes a superpopulation point of view in the terminology of~\cite{imbens2015causal}.   Results from martingale theory \citep{fleming2013counting,abadie2012martingale} and empirical process theory \citep{van1996weak} are used to study the asymptotic behavior of generalized coarsened confounding-based estimators.   While our results are relatively general, this paper develops a formal theoretical justification for the variance estimates for causal effects from coarsened exact matching of~\cite{iacus2011multivariate}.   In addition, we develop two new algorithms for generalized coarsened confounding using machine learning.  The first is an adaptation of the k-means algorithm~\citep{macqueen1967some} for estimating causal effects. The second is based on random forests~\citep{breiman2001random}.  

One implication of our analysis in Section ~\ref{sec:methods} is that there will be a bias in our estimates in finite samples.   This is because the creation of discrete strata will induce an estimation bias.  While a similar phenomenon was observed for nearest neighbor matching-based causal effect estimation by  \citet{abadie2006large}, we cannot use their techniques for analysis and bias correction.  This necessitates a new approach to bias correction that we discuss in Section~\ref{sec:biasc}. The structure of this paper is as follows.   In Section~\ref{sec:background}, we set up notation, assumptions and review related literature.   Section~\ref{sec:methods} describes the coarsened confounding paradigm and develops some results for asymptotic results for coarsened exact matching as well as the new k-means algorithm for causal effect estimation. Section~\ref{sec:examples} shows some evaluations using real data examples for the causal effect estimators.  Section~\ref{sec:discussion} concludes with some discussion.

%% file: sections/background.tex
\subsection{Data Structures and Causal Estimands}

Let the data be represented as $(Y_i,T_i,{\bf X}_i)$, $i=1,\ldots,n$, a
random sample from the triple $(Y,T,{\bf X})$, where $Y$
denotes the response of interest, $T$ denotes the treatment group, and ${\bf X}$ is a $p$-dimensional vector of covariates. We assume that
$T$ takes values {in} $\{0,1\}$.  

We now briefly review the potential 
outcomes framework of \citep{rubin1974estimating} and \cite{holland1986statistics}.  
Let $\{Y(0),Y(1)\}$ denote the potential outcomes for all $n$ subjects, and
the observed response {be} related to the potential outcomes by
\[Y = (1-T)Y(0) + TY(1).\]
In the potential outcomes framework, causal effects are defined as within-individual contrasts based on the potential outcomes.  
One popularly used estimand is the average causal effect, defined as 
\[
\text{ACE} = \frac{1}{n} \sum_{i=1}^n \left(Y_i(1) - Y_i(0)\right).
\]

Many assumptions are needed for performing valid causal inference.  These include the consistency assumption, the treatment positivity assumption, and the strongly ignorable treatment assumption \citep{rosenbaum1983central}.  Consistency means that the potential outcome for the observed treatment and the observed outcome are the same.    Treatment positivity refers to $0 < P(T = 1\mid {\bf X}) < 1$ for all values of ${\bf X}$. The intuitive interpretation of the positivity assumption is that any individual can potentially receive either treatment, although its validity for high-dimensional ${\bf X}$ has been recently questioned by \citet{d2021overlap} and \citet{ghosh2019gaussian}.

The strongly ignorable treatment assumption is defined as 
\begin{equation}\label{unc}
T \bigCI  \{Y(0),Y(1) \} \mid {\bf X}.
\end{equation}
Assumption (\ref{unc}) means that treatment assignment is conditionally independent of the set of potential outcomes given the covariates. Finally, the consistency assumption ensures that the observed outcome and the potential outcome under the observed treatment coincide.    

As described recently by \citet{imbens2015causal}, causal inference proceeds by modelling the assignment mechanism using observed covariates.   A quantity that naturally arises from this modelling is the propensity score \citep{rosenbaum1983central}, the probability of receiving treatment given confounders.  
The propensity score is defined as
\[
e({\bf X}) = P(T = 1\mid {\bf X}).
\]
Given the treatment ignorability
assumption in (\ref{unc}), it also follows by Theorem 3 of \citet{rosenbaum1983central} that treatment is
strongly ignorable given the propensity score, i.e.
\[ {T} \bigCI 
\{Y(0),Y(1)\} \mid e({\bf X}).\] 

Based on these assumptions and definitions, we can formulate causal inference using the following approach: (a) define an appropriate causal estimand; (b) formulate a propensity score model;
(c) check for covariate balance; (d) if (c) holds, estimate the causal estimand by conditioning on the propensity scores. We note that steps (b) and (c) tend to be iterative in practice. 


\subsection{Coarsened Exact Matching}\label{sec:cem}


\citet{iacus2011multivariate} took another approach to causal inference by focusing on in-sample covariate discrepancies and requiring that the maximum discrepancy in sample means between treated and control subjects be bounded above by a constant.  They generalize this to arbitrary functions of the data, which they term imbalance bounding and define monotonic imbalance bounding matching methods to be those in which the discrepancies between a monotonic function applied to a variable is bounded above by a confounder-specific term.
Thus, one can be more stringent in the balance in variables without impacting the maximal imbalance across all confounders.  

There are many important implications of requiring the monotonic imbalance bounding property.  First, many methods  of confounder adjustment, such as nearest-neighbor or caliper matching as defined in \cite{cochran1973controlling}, are not monotonic imbalance bounding because they fix the number of treated and control observations within strata, while monotonic imbalance bounding methods imply variable numbers of observations.  By contrast, if the caliper matching procedure were to allow for different calipers for each confounder, then this would be monotonic imbalance bounding. \citet{iacus2011multivariate} also show that a key goal in causal effect estimation is to reduce model dependence \citep{ho2007matching}, meaning that there should not be  extrapolation of potential outcomes to regions in the covariate space where there are no observations. Under some assumptions on the model for potential outcomes, they show that for monotonic imbalance bounding methods, the model dependence is upper bounded by terms involving an imbalance parameter.
In addition, the estimation error for average causal effects using monotonic imbalance bounding matching methods can also be upper bounded by terms involving this parameter.

As a concrete example of a new monotonic imbalance bounding method, \citet{iacus2011multivariate} propose coarsened exact matching for creating strata.  It proceeds as follows:
\begin{enumerate}
\item For each component of ${\bf X}$, $X_j$ $(j=1,\ldots,p)$, coarsen it into a function $C_j(X_j)$ which takes on fewer values than the unique values of $X_j$;
\item Perform exact matching between treated and control observations using the vector 
\[\left(C_1(X_1),C_2(X_2),\ldots,C_p(X_p)\right).\]
This effectively creates strata ${\cal S}_1,\ldots,{\cal S}_J$ based on the unique combinations of 
\[\left(C_1(X_1),C_2(X_2),\ldots,C_p(X_p)\right).\]
\item Discard strata in which there are only observations with $T = 0$.  For strata with only observations from the $T = 1$ population, extrapolate the potential outcome $Y(0)$ using the available controls or discard by restricting the causal effect of interest on the treated units for which causal effect can be identified without further modelling based assumptions. For strata with both treated and control observations, compare the outcome between the two populations.   
\end{enumerate}
\citet{iacus2011multivariate} have developed very easy-to-use software packages for implementing coarsened exact matching in R and Stata. They show that the coarsened exact matching approach satisfies the monotonic imbalance bounding property with respect to a variety of functionals of interest.  In addition, they provide a very intuitive explanation for what coarsened exact matching attempts to mimic. While classical propensity score approaches attempt to mimic a randomized study, analyses using coarsened exact matching will mimic randomized block designs, where the blocks are by definition predictive of the potential outcomes.  It is well-known that in this situation, randomized block designs will yield more efficient estimators \citep{box1978statistics}.

\subsection{Related literature}\label{sec:previous}

One related apporach to coarsened exact matching is subclassification.  There has been limited exploration on the use of propensity score subclassification from \cite{cochran1968effectiveness} and \cite{cochran1973controlling}.   \cite{hullsiek2002propensity} explored the issue of the construction of strata to use for propensity score subclassification.  They proposed an algorithm for strata construction.  It involved creating initial sets of equally sized strata and then to iteratively adjust the sizes based on the estimate variance of the treatment effect. The simulation studies in \cite{hullsiek2002propensity} confirmed the findings of \citet{cochran1968effectiveness} under a generative linear propensity score model framework but also stressed that simply having subclasses of the same size did not necessarily guarantee stable causal effect estimators.  We will discuss the applicability of our results to propensity score classification in Section~\ref{sec:asymp}.

Another related literature is the theory of  {\it equal percent bias reduction} procedures \citep{rubin1976multivariate,rubin1992affinely,rubin2000combining,rubin2006affinely}.  Equal percent bias reduction means that a certain type of covariate matching will reduce bias in all dimensions of ${\bf X}$ by the same amount.
We define a matching method to be affinely invariant if the matching procedure is invariant to affine transformations of the covariates. If ${\bf X}$ given $T$ is assumed to have an elliptically symmetric distribution, then Theorem 3.1. and Corollaries 3.1. and 3.2 of \citet{rubin1992affinely} apply so that any affinely invariant matching method will be equal percent bias reducing.   Examples of elliptically symmetric distributions include the multivariate normal and t distributions.   While elliptical symmetry of the confounders given treatment group is a restrictive assumption, this was relaxed in more recent work by \citet{rubin2006affinely}. There, they assumed that the conditional distribution of ${\bf X}$ given $T$ is a discriminant mixture of elliptically symmetric distributions. \citet{rubin2006affinely} prove that a generalization of equal percent bias reducing holds for this setup as well. 
As mentioned in \citet{iacus2011multivariate}, while equal percent bias reduction represents a superpopulation property, the monotonic imbalance property is an in-sample property that the authors is more general.  

Finally, there is a rich literature on matching methods that this paper adds to.  Matching algorithms are a set of procedures that attempt to find for each treated observation in the dataset, the control observation that is `closest' in terms of confounder values.   There are many methods available for matching, including nearest-neighbor matching, K:1 matching and optimal matching~\citep{rosenbaum1989optimal}.   Once the matched sets are constructed, a variety of approaches can be used to estimate the causal effected in the matched dataset.  A nice summary of methods for estimating causal effects in matched datasets can be found in Section 5 of~\citet{stuart2010matching}.  Proposed solutions include regression adjustments, weighted analysis approaches and hybrid combinations thereof.     

\cite{abadie2006large} studied the theoretical properties and asymptotics of nearest-neighbor matching procedures.  They followed this with work studying the asymptotics of matching on the estimated propensity score~\citep{abadie2016matching}.   A difference between coarsened exact matching with nearest-neighbor matching is that the former does not use any information on $T$ or $Y$ in order to generate the random sets.  By contrast, for nearest-neighbor matching, one finds the `closest' control for every treated observation, where the distance is defined based on an appropriate chosen metric for the confounders. \cite{abadie2006large} used Euclidean distance assuming the confounders were all continuous.  They made the following observations: 
	\begin{enumerate}
		\item There will be an asymptotic non-negligible bias in the estimate of the average causal effect that is a function of the amount of covariate imbalance within strata;
		\item This bias correction will have to be estimated from the data, but provided it can be estimated reliably, the resulting bias-adjusted inference will be asymptotically valid.
	\end{enumerate}
As we discuss in \S~\ref{sec:asymp}, there is also a finite-sample bias to coarsened confounding; however, we will be unable to use the approach in \citet{abadie2011bias}.


%% file: sections/methods.tex
\subsection{Review and Structure Identification}
\label{sec:methods}
Using the notation from Section~\ref{sec:background}, we can express the coarsened exact matching causal effect estimator as 
\begin{equation}\label{taucem}
\widehat \tau_{CEM} \equiv \sum_{j=1}^J \frac{n_j}{n}(\bar Y_{1j} - \bar Y_{0j}).
\end{equation}
In (\ref{taucem}), $\bar Y_{1j}$ and $\bar Y_{0j}$ denote the sample averages for the response in the $j$th stratum, $j=1,\ldots,J$.   \cite{iacus2011multivariate} suggested the following variance estimate for $\widehat \tau_{CEM}:$

\begin{equation}\label{vartaucem}
\widehat \sigma^2_{CEM} = \sum_{j=1}^J \left (\frac{n_j}{n}\right)^2 \left (\frac{\widehat \sigma^2_{1j}}{n_{1j}} + \frac{\widehat \sigma^2_{0j}}{n_{0j}} \right ),
\end{equation}
where $\widehat \sigma^2_{1j}$ and $\widehat \sigma^2_{0j}$ denote the estimated variances of $Y$ in the treated and control groups for the $j$th stratum.   The quantities $n_{0j}$ and $n_{1j}$ represent the number of control and treated observations, respectively, in the $j$th stratum. 

The reinterpretation that we will heavily leverage in this article is that coarsened exact matching generates strata ${\cal S}_1,\ldots,{\cal S}_J$ using ${\bf X}_1,\ldots,{\bf X}_n$.   We can rewrite the average causal effect $\widehat \tau_{CEM}$ from (\ref{taucem}) as solving the estimating equation $U(\tau) = 0$, where
	\begin{equation}\label{ee1} 
		U(\tau) =  \sum_{i=1}^n  \sum_{j=1}^J I(i \in {\cal S}_j) \frac{n_j}{n} \left [ \frac{T_iY_i}{n_{1j}} - \frac{(1-T_i)Y_i}{n_{0j}}  - \frac{\tau}{nJ} \right ] 
	\end{equation}
	The parameter $\tau$ in (\ref{ee1}) represents the population average causal effect.   
 If we were to condition on the ${\cal S}_j$'s in (\ref{ee1}), then the only randomness is in $Y$ so that we can proceed using standard estimating function arguments~\citep{tsiatis2006semiparametric}.  However, the more general case allows for randomness in ${\cal S} \equiv ({\cal S}_1,\ldots,{\cal S}_J)$, which is what weconsider here.    Thus, one can interpret (\ref{ee1}) as a random set-induced estimating equation. 

We note the centrality of equation (\ref{ee1}) to our generalized coarsened confounding framework.   Provided we have an algorithm ${\cal M}$ that takes as its input  ${\bf X}_1,\ldots,{\bf X}_n$ and provides as output the strata ${\cal S}_1,\ldots,{\cal S}_J$, we can use (\ref{ee1}) to compute the average causal effect.    While~\citet{iacus2009cem}
 use hypercubes based on the components of ${\bf X}$ to define strata, in practice any algorithm for clustering confounders could be used.  We will study two other algorithms in this paper, k-means and random forests clustering.   What also is evident from (\ref{ee1}) is that the asymptotic behavior of the estimator will necessitate studying the asymptotic behavior of the sets ${\cal S} \equiv ({\cal S}_1,\ldots,{\cal S}_J)$, and this is what we do in Section~\ref{sec:asymp}.  We next describe two machine-learning based algorithms for clustering.  
 
 \subsection{A quantization-based causal effects estimator using k-means}
\label{sec:kmeans}

For ease of exposition, we assume that each of the $p$ confounders are coarsened into $M$ levels.   Thus, there are $M^p$ possible values for the levels.  In the terminology of information theory
~\citep{wolfowitz2012coding}, each of these levels constitutes a code, and the set of potential values is called the codebook.  In coarsened exact matching, a codebook with $M^p$ possible values is constructed.  The values constitute strata in which responses for treated and controls are compared.     Thus, the problem of causal effect estimation is effectively reduced to one of constructing codebooks, which has been a foundational topic in information theory for decades, dating back to the work of \citet{shannon1948mathematical}.   Thus, the random sets ${\cal S}_1,\ldots,{\cal S}_J$ can be viewed as codebook representations for the confounders.   We term this confounder representation learning, which is a specialized version of a more general phenomenon, representation learning, that has received tremendous attention in the machine learning community~\citep{bengio2013representation}.

We have argued that coarsened exact matching refers to one type of codebook construction.  Another term for this is quantization; the ${\cal S}_j$'s denote the quantized levels of the confounders.   The problem of quantization has a long rich history in information theory~\citep{gray1984vector,graf2007foundations,gersho2012vector}.  In this area, a workhorse technique for quantization has been the $k-$means algorithm \citep{macqueen1967some,lloyd1982least}, which we next describe.   

Let ${\bf X}$ denote the $p-$dimensional vector of confounders, which we assume to have marginal distribution $P_{\bf X}$.  Define $\|a\|$ to be the norm of $a$ and assume that $E\|{\bf X}\|^2 < \infty.$ We then define a center of $P_{\bf X}$ to be a point ${\bf b} \in R^p$ such that
\begin{equation}\label{center}
    E\|{\bf X} - {\bf b}\|^2 = \inf_{{\bf a} \in R^p} E\|{\bf X} - {\bf a}\|^2.
\end{equation}
It turns out that the solution to (\ref{center}) has an equivalent interpretation in terms of function approximation.   Let ${\cal F}_m$ be the set of Borel-measurable functions $f: R^p \rightarrow R$ with $|f(R^p)| \leq m$.   The elements of ${\cal F}_m$ are termed $m-$point quantizers.  The $m-$quantization error for $P_{\bf X}$ of order two is defined by 
\begin{equation}\label{vnr}
V_{m}(P_{\bf X}) = \inf_{f \in {\cal F}_m} E\|{\bf X} - f({\bf X})\|^2.
\end{equation}
A quantizer $f \in {\cal F}_m$ is called $m-$optimal for $P_{\bf X}$  if 
\begin{equation}\label{vnropt}
V_{m}(P_{\bf X}) =  E\|{\bf X} - f({\bf X})\|^2.
\end{equation}
A set ${\bf A} \subset R^p$ with $|{\bf A}| \leq m$ and where
$$ E \min_{{\bf a} \in A} \|{\bf X}-{\bf a}\|^2 = \inf_{{\bf A} \subset R^p ,|{\bf A}| \leq m} E\min_{{\bf a} \in {\bf A}} \|{\bf X} - {\bf a}\|^2$$
is called an $m-$optimal point set of centers of $P_{\bf X}$.
From Lemma 3.1. of \cite{graf2007foundations}, we have that 
\begin{equation}\label{opt1}
V_{m}(P_{\bf X}) = \inf_{ {\bf A} \subset {\bf R}^p,
|{\bf A}| \leq m} E\min_{{\bf a} \in {\bf A}} \|{\bf X}-{\bf a}\|^2.
\end{equation}
In practice, the empirical version of quantizers has an intimate connection with the $k-$means algorithm~\citep{macqueen1967some,lloyd1982least}.  The basic k-means algorithm proceeds in the following steps: 
\begin{itemize}
    \item[(a).] assign an initial set of means $\tilde {\bf x}_1,\ldots,\tilde {\bf x}_K$ in $R^p$; 
    \item[(b).] assign each observation to the cluster with the nearest mean; 
    \item[(c).]
    recalculate the means for the observations assigned to the cluster; 
    \item[(d).] iterate between steps b. and c. until convergence.  
\end{itemize}
While the k-means algorithm is known to statisticians as primarily a clustering technique, it plays a central role in the quantization literature in information theory.   Under the assumption of the existence of a unique population maximizer, \cite{pollard1981strong} demonstrated the convergence of the optimizer of the k-means algorithm to the population limit using empirical process theory.  Arguments of the convergence using fixed point theory are provided in \cite{kieffer1982exponential} and in \cite{sabin1986global}.  

The k-means algorithm yields a set of clusters that constitute the strata within which we can compare the outcomes between treated and controls, exactly as in \cite{iacus2011multivariate}.  We thus have the following simple algorithm for causal effect estimation.  
\begin{algorithm}\label{algo1}
    \caption{Proposed confounder adjustment algorithm using k-means clustering}
    \begin{enumerate}
 
\item Cluster confounders using $k-$means, which generates $k$ strata;
\item Compute the causal effect by comparing $Y$ between the treated and control groups within the strata;
\item Compute the variance of the average causal effect using Equation~\ref{vartaucem}.  Thus, the output will be an estimated 
causal effect estimate and associated confidence interval.
       
    \end{enumerate}
\end{algorithm}
\noindent Notice that Algorithm~\ref{algo1} is identical to coarsened exact matching with exception of Step 1. Geometrically, the CEM is constructing hypercubes for strata, whereas our approach creates ellipses for the confounder levels. 

In information theory, a key quantity is the rate distortion function~\citep{berger2003rate}.  It summarizes how much information can be preserved using data compression methods.   It corresponds to the entropy change after data compression.   Coarsened exact matching corresponds to taking confounders and recoding it as binary indicators indexing the hypercubes.  Assuming $M$ levels for each of the $p$ confounders, the data compression, is approximately $p \log M$. By contrast, $k-$means takes the $n$ observations and maps them to $k$ cluster means, which suggests that the data compression is effectively $\log k$.  Note that the data compression for $k-$means clustering is independent of dimension of the confounders.  By contrast, the data compression scales linearly in the number of confounders for coarsened exact matching.  In information theory, a major goal is to devise procedures that maximize information compression while at the same time preserving information.  With respect to the former, our k-means algorithm offers an advantage relative to coarsened exact matching.  

\subsection{Random forests}
While the use of k-means can potentially have theoretical advantages relative to coarsened exact matching, it suffers from several drawbacks.   First, it is most effective when the confounders are continuous, and the performance might degrade when there are categorical variables.   Second, the clusters are assumed to elliptical in nature, which may or may not be a reasonable assumption.  Third, there is an equivalence between k-means clustering with maximum likelihood of a specific normal mixture model~\citep{fraley2002model}, there is a linearity of the variable information that is used in the algorithm, which may or may not hold.   In order to allow for more flexibility in strata construction and more easily accommodate mixed continuous and discrete variables, we propose to use random forests~\citep{breiman2001random}.

Random forests represent a class of ensemble methods: instead of generating one classification tree, it generates many trees. At each node of a tree, a random subset of the covariates are selected and the node is split based on the best split among the selected covariates. For a testing data point with a covariate vector {\bf X}, each tree votes for one of the classes and the prediction can be made  by the majority votes among the trees. In addition, some appealing by products of random forests include the following: (a) a variable importance measure; (b) an out of bag estimator of the model performance; (c) a measure of observational proximity.    Random forests represent among the most popular off-the-shelf machine learning methods and require minimal amounts of tuning. While they are popular, theoretical justification for their use is an area of intense focus~\citep{biau2008consistency,biau2012analysis,biau2016random}.  

Much of the previous work on random forests assumes a supervised setting in which there is an outcome variable.   We wish to use an unsupervised version of random forests.  To do this, we adopt what was suggested in ~\citet{breiman2001random}, which is to treat the observed data as coming from one class and to create synthetic data that from a second group.  Then random forests classification is performed on the augmented dataset.  We then take the so-called proximity matrix and cluster observations into strata using Ward's method of clustering~\citep{ward1963hierarchical}.   This generates a dendrogram representing a hierarchical grouping; we will cut the dendrogram at a certain level to create strata.   This replaces Step 1 of Algorithm~\ref{algo1}.

\subsection{Asymptotic Analysis}\label{sec:asymp}
Note that (\ref{taucem}) and (\ref{vartaucem}) can be generalized to allow for more general constructions of strata ${\cal S}_j$, $j=1,\ldots,J$, which we reexpress as

\begin{equation}\label{taus}
\widehat \tau_{S} \equiv \sum_{{\cal S}_j,j=1,\ldots,J} \frac{n_j}{n}(\bar Y_{1j} - \bar Y_{0j}).
\end{equation}
with estimated variance 
\begin{equation}\label{vartaus}
\widehat \sigma^2_{S} = 
\sum_{{\cal S}_j,j=1,\ldots,J} \left (\frac{n_j}{n} \right )^2 \left (\frac{\widehat \sigma^2_{1j}}{n_{1j}} + \frac{\widehat \sigma^2_{0j}}{n_{0j}} \right ),
\end{equation}
Thus, formulae (\ref{taus}) and (\ref{vartaus}) allow for CEM, the k-means estimator and random forests-based clustering with the attendant standard errors.   More generally, we can allow for ${\cal S}_j$ $(j=1,\ldots,J)$ to be based on any data-driven algorithm for partitioning based on ${\bf X}$.  However, it cannot be based on $T$ or ${\bf Y}$. Having done this, the next question is to understand the asymptotic properties of the estimators.   
We note the following decomposition for $\widehat \tau_S - \tau$:
\begin{eqnarray*}
\widehat \tau_S - \tau &=& n^{-1} \sum_{i=1}^n \left [\mu_1({\bf X}_i) - \mu_0({\bf X}_i) - \tau \right ] \\
&&+ n^{-1} \sum_{i=1}^n \left \{ \left [ \sum_{j=1}^J \frac{n_j}{n} I(i \in {\cal S}_j) \frac{T_iY_i}{n_{1j}/n}  \right ]- \mu_{1}({\bf X}_i) \right \} \\
&&-  n^{-1} \sum_{i=1}^n \left  \{ \left [ \sum_{j=1}^J \frac{n_j}{n} I(i \in {\cal S}_j) \frac{(1-T_i)Y_i}{n_{0j}/n}  \right ]- \mu_{0}({\bf X}_i) \right \}. 
\end{eqnarray*}
Let ${\cal X}_n = \{{\bf X}_1,\ldots,{\bf X}_n\}$ and ${\cal T}_n = \{T_1,\ldots,T_n\}$.  Next, we define the random variables
\begin{equation*}
\xi_{n,k} = \begin{cases} n^{-1/2} (\mu_1({\bf X}_k) - \mu_0({\bf X}_k) - \tau), \ \ \ 1 \leq k \leq n \\
\\
n^{-1/2}  \left \{ \left [ \sum_{j=1}^J \frac{n_j}{n} I(k \in {\cal S}_j) \frac{T_kY_k}{n_{1j}/n}  \right ]- \mu_{1}({\bf X}_k) \right \} - 
n^{-1/2} \left \{ \left [ \sum_{j=1}^J \frac{n_j}{n} I(k \in {\cal S}_j) \frac{T_kY_k}{n_{0j}/n}  \right ]- \mu_{0}({\bf X}_k) \right \}. \\
\ \ \ \ n+1 \leq k \leq 2n\\
\end{cases}
\end{equation*}
We can then write
\begin{eqnarray}\label{decomp}
\sqrt{n}(\hat \tau_S - \tau) &=& 
 \sum_{k=1}^n \xi_{n,k} + \sum_{k=n+1}^{2n} \xi_{n,k}  \nonumber\\
&=& W_n + R_{n}, 
\end{eqnarray}
where $W_n$ can be interpreted as a normalized average of mean zero iid random variables, and $R_{n}$ represent the conditional bias terms due to the strata creation and comparing the average outcomes to the potential outcome functions for the treatment and control groups, respectively. 
Finally, define the $\sigma-$field
\begin{equation*}
{\cal F}_{n,k} = \begin{cases} \sigma\{{\cal T}_n,{\bf X}_1,\ldots,{\bf X}_k\}, \ \ \ \  k=1,\ldots,n \\
\\
\sigma\{{\cal T}_n,{\cal X}_n,Y_1,\ldots,Y_{k-n}\}, \ \ \ n+1 \leq k \leq 2n.
\end{cases}
\end{equation*}
Several facts obtain from (\ref{decomp}).  First, 
as in \cite{abadie2012martingale}, 
$$ \left \{ \sum_{j=1}^i \xi_{n,j}, {\cal F}_{n,i}, \ \ \ 1 \leq i \leq 2n \right \}$$
represents a martingale for each $n \geq 1$.  Equivalently, the collection represents a martingale array.    Thus, we can use results from martingale theory~\citep{fleming2013counting} to study the asymptotics of $\sqrt{n}(\hat \tau_S - \tau)$.  Second, there is a nonneglible bias term, $R_n$, 
that represents how well a piecewise constant function can approximate the potential outcome functions.  As is well-known in nonparametric theory~\citep{devroye2013probabilistic}, for a finite number of strata ${\cal S}_j$ $(j=1,\ldots,J)$, for $t=0,1$, $\mu_t(\cdot)$ will not be consistently estimated.  Put another way, we need the number of strata $J_n$ to approach infinity as the sample size tends to infinity as well. 
This implies that one ought to use a large number of strata.  However, we find that as the number of strata gets bigger, we will have strata which contain only treated or control observations.  We must then exclude those strata, which reduces the effective sample size in our analysis.  

We now prove this more formally by leveraging results from Chapters 12 and 21 of \cite{devroye2013probabilistic}.    Assume that the joint distribution of the confounders ${\bf X}$ has a measure $\mu$ on $R^p$.  A partition of $R^p$ is a countable collection of sets $\{{\cal A}_n, n \geq 1\}$ such that $\cup_{n=0}^{\infty} {\cal A}_n = R^p$ and ${\cal A}_j \cap {\cal A}_k = \emptyset$ for $j \neq k$.  We refer to each ${\cal A}_j$ as a cell.   Define $M > 0$, and take $S_M \subseteq R^p$ to be the closed ball of radius $M$ centered at the origin.   For each partition, we define ${\cal P}^{(M)}$ as the restriction of ${\cal P}$ to $S_M$.  Define ${\cal B}({\cal P}^{(M)})$ to be the collection of all $2^{|{\cal P}^{(M)}|}$ sets formed by taking unions of cells in ${\cal P}^{(M)}.$   Let ${\cal G}$ be a potentially infinite collection of partitions of $R^p$; define ${\cal G}^{(M)} = \{ {\cal P}^{(M)}: {\cal P} \in {\cal G}\}$ to be the family of partitions of $S_M$ obtained by restricting ${\cal G}$ to $S_M$.   Define ${\cal C}^{(M)}$ to be the class of subsets of $R^p$ where
$${\cal C}^{(M)} = \{ A \in {\cal B}({\cal P}^{(M)}) \text{ for some }
{\cal P}^{(M)} \in {\cal G}^{(M)}\}.$$
For $(z_1,\ldots,z_n) \in \{R^p\}^n$, let  
${\cal N}_{{\cal C}^{(M)}}(z_1,\ldots,z_n)$ be the number of different sets in 
$$ \{ \{z_1,\ldots,z_n\} \cap C: C \in {\cal C}^{(M)}\}.$$
The $n-$th shatter coefficient of ${\cal C}^{(M)}$ is
\begin{equation}\label{shattercoef}
s({\cal C}^{(M)},n) = \max_{\{z_1,\ldots,z_n\} \in \{R^p\}^n}  {\cal N}_{{\cal C}^{(M)}}(z_1,\ldots,z_n);
\end{equation}
in words, (\ref{shattercoef}) represents the maximal number of different subsets of $n$ points that can be picked out by the class of sets ${\cal C}^{(M)}$.   We define the following combinatorial quantity on the partitions ${\cal G}^{(M)}$:
$$\Delta_n({\cal G}^{(M)}) = s({\cal C}^{(M)}, n).$$
The combinatorial quantity $\Delta_n({\cal G}^{(M)})$ represents the complexity of the partitions as the sample size increases. We let $\mu_n$ denote the empirical measure of ${\bf X}_1,\ldots,{\bf X}_n$.   We then have the following result from \citet{lugosi1996consistency}:

\noindent {\bf Theorem 1:}\citep{lugosi1996consistency} Let ${\bf X}_1,\ldots,{\bf X}_n$ be iid random vectors in $R^p$ with measure $\mu$ and empirical measure $\mu_n$.  Let ${\cal G}$ be a collection of partitions on $R^p$.    Then for each $M < \infty$ and $\epsilon > 0$,

\begin{equation}\label{lnconc}
P\left \{ \sup_{{\cal P}^{(M)} \in {\cal G}^{(M)}} \sum_{A \in {\cal P}^{(M)}} |\mu_n(A) - \mu(A)| > \epsilon \right \} \leq 8\Delta_n({\cal G}^{(M)})\exp(-n\epsilon^2/512) + \exp(-n\epsilon^2/2).
\end{equation}
An implicit assumption in (\ref{lnconc}) is that appropriate measurability conditions on the probability on the left-hand side of the inequality is needed.  Conditions to ensure the measurability of the supremum of events can be found in \citet{van1996weak}.  Leveraging Theorem 1, we prove the following result.
\\
\\
\noindent {\bf Theorem 2:}  Assume that there exist a sequence of families ${\cal G}^{(M)}_n$ $(n \geq 0)$ such that
\begin{equation}\label{asympgrowth} 
\lim_{n \rightarrow \infty} \frac{\log \Delta_n({\cal G}_n^{(M)})}{n} = 0.
\end{equation}
Then 
\begin{equation}
n^{1/2}(\tau_S - \tau) \rightarrow_d N(0,\sigma^2),
\end{equation}
where $\rightarrow_d$ denotes convergence in distribution, and 
$$ \sigma^2 = E[(\mu_1({\bf X}) - \mu_0({\bf X}) - \tau)^2].$$

\noindent {\bf Proof:}  Based on (\ref{decomp}), 

\begin{eqnarray}\label{decomp2}
\sqrt{n}(\hat \tau_S - \tau) &=& W_n + R_{n} 
\end{eqnarray}
In (\ref{decomp2}), $W_n$ converges in distribution to a normal random variable with mean zero and variance $\sigma^2$ by the martingale central limit theorem~\citep{fleming2013counting}.   To deal with $R_n$, we use Theorem 1 and the assumption (\ref{asympgrowth}) to show that $R_n \rightarrow 0$ in probability.  Application of Slutsky's theorem then yields the desired result.

\noindent {\bf Remark 1:}  In Theorem 2, we also provide a sufficient condition for the bias term in (\ref{decomp}) to be asymptotically negligible.   We note that condition (\ref{asympgrowth}) in Theorem 2 is in fact an asymptotic condition on the complexity of the sets that are induced by the partitioning algorithm.    We now discuss the implications of the condition for coarsened exact matching, k-means and random forests based clustering.   For coarsened exact matching, we assume that for each of the $p$ covariates, we partition them into $M$ bins.   Some standard combinatorial arguments (e.g., p. 367 of ~\citet{devroye2013probabilistic}) can be used to show that for this set of partitions, ${\cal G}_n^{(M)}$,
$$ \Delta_n({\cal G}_n^{(M)})  \leq 2^{M^p} \binom{n+M}{n}^{p}.$$  We thus have that (\ref{asympgrowth}) is satisfied when 
$$ \lim_{n \rightarrow \infty} \frac{M_n^{p}}{n} = 0.$$
For the $k-$means partition approach to clustering, we leverage the work of~\citet{lugosi1996consistency}, who demonstrate concentration bounds for Voronoi partitions of points in $R^p$.   Based on our notation and their results, defining $J_{kn}$ to be the closest neighbors of the $k$th cluster center and allowing for dependence on sample size, if 
$$ J_{kn} \rightarrow \infty \ \ \text{and} \ \  \frac{J^2_{kn} \log n}{n} \rightarrow 0$$
as $n \rightarrow \infty$, then (\ref{asympgrowth}) is satisfied.  Finally, for the the random forests approach, we begin by assuming a single tree with at most $S_n$ splits, where 
$$ \frac{S_n \log n}{n} \rightarrow 0.$$
as sample size approaches infinity.   Then the results of~\citet{lugosi1996consistency} can again be used to verify that (\ref{asympgrowth}) holds.    Random forests now aggregates over $B$ trees, so the condition on $S_n$ will again imply that for the aggregated tree, (\ref{asympgrowth}) holds.   We wish to reiterate that condition (\ref{asympgrowth}) holding is theoretical and relies on asymptotic considerations in which sample size and other tuning parameters are approaching infinity with potential constraints on their interactions.   

Theorem 2 provides a formal justification of the variance estimators proposed by~\citet{iacus2011multivariate} for coarsened exact matching.  Note that our theorem is much more general and shows the asymptotic normality for the various data-partitioned causal effect estimators we have proposed: coarsened exact matching, k-means clustering, and random forests clustering.     We note in passing that an alternative to using empirical process theory results to characterize the nature of the bias term in (\ref{decomp}) would be to use sample splitting/cross-fitting~\citep{zivich2021machine} to estimate the bias term directly.   Many authors (e.g.,~\citealp{chernozhukov2018double}) have shown that based on sample splitting, one can weaken the smoothness conditions necessary for the bias term to converge to zero in probability.   We do not pursue that here.   

We now seek to mention some other implications of the martingale theory results we have presented here.   First, the construction of the filtrations and $\sigma-$algebras in the theory means that rules for creating strata that are based on ${\bf X}$ only will have the same theoretical properties as those demonstrated here.  We described propensity score subclassification as a related methodology in \S~\ref{sec:previous}.   Our theory would apply to subclassification estimators with a {\bf known} propensity score.  In this case, strata construction would proceed based on $e({\bf X}) = P(T = 1|{\bf X})$, which is simply a function of ${\bf X}$.  However, if we were to fit a model for the propensity score to data, then constructing strata based on $\widehat e({\bf X})$ would not fall under our framework.  While the notion of using population or known propensity scores do not seem reasonable in practice, we note in passing that much of the work in the seminal paper of~\cite{rosenbaum1983central} effectively works with $e({\bf X})$ and not $\widehat e({\bf X})$.    
Finally, we do wish to point out that this approach to causal effect estimation avoids propensity scores, but in effect does not model $T|{\bf X}$ directly.   Thus, it represents a very different theoretical framework than most causal effect-based estimators, which use estimating functions and semiparametric theory results~\citep{tsiatis2006semiparametric}.   However, we feel there is merit in understanding the theoretical basis for generalized coarsened confounding, and we believe there are extensions of the approach that could handle more complicated confounding structures, which we leave to future investigations.  

\subsection{Average Causal Effect on the Treated}\label{sec:att}
An alternative estimand in causal inference is the average causal effect among the treated:
\begin{equation}\label{acet}
\tau_A = E[Y(1)-Y(0)|T=1].
\end{equation}
ACET is of particular interest when the population of the study are those who actually receive the treatment. For example, a researcher from a smoking cessation counseling tries to persuade the smokers to quit smoking and his research question is as follows: for those who actually smoke, what is the difference in the expected life expectancy if they did not smoke? In this example, the researcher is interested in estimating ACET.  This is in fact the primary estimand studied by ~\cite{iacus2011multivariate} for coarsened exact matching.   Their estimators are
\begin{equation}\label{tausacet}
\widehat \tau_{S,A} \equiv \sum_{{\cal S}_j,j=1,\ldots,J} \frac{n_j}{n} (\bar Y_{1j} - w_j \bar Y_{0j}).
\end{equation}
with estimated variance 
\begin{equation}\label{vartausacet}
\widehat \sigma^2_{S,A} = 
\sum_{{\cal S}_j,j=1,\ldots,J} \left (\frac{n_j}{n}\right )^2 \left (\frac{\widehat \sigma^2_{1j}}{n_{1j}} + w^2_j \frac{\widehat \sigma^2_{0j}}{n_{0j}} \right ),
\end{equation}
where 
$$ w_j = \frac{n_{1j}/n_1}{n_{0j}/n_0}.$$
We can then derive a result similar to Theorem 2 using the following decomposition:
\begin{eqnarray*}
\widehat \tau_{S,A} - \tau_A &=& n^{-1} \sum_{i=1}^n T_i \left [\mu_1({\bf X}_i) - \mu_0({\bf X}_i) - \tau_A \right ] \\
&&+ n^{-1} \sum_{i=1}^n  T_i \left   \{ \left  [ \sum_{j=1}^J I(i \in {\cal S}_j) \frac{T_iY_i}{n_{1j}/n}  \right ]- \mu_{1i}({\bf X}_i) \right \} \\
&&+ n^{-1} \sum_{i=1}^n T_i \left  \{ \left [ \sum_{j=1}^J I(i \in {\cal S}_j) \frac{(1-T_i)Y_i}{n_{0j}/n}  \right ]- \mu_{0i}({\bf X}_i) \right \}. 
\end{eqnarray*}

\subsection{Bias correction}\label{sec:biasc}

As alluded by Theorem 2,  provided the number of strata increases with sample size in such a way so that the complexity of the partitions grows slowly, the coarsened causal effects estimator will be consistent.  However, in practice, the number of strata is finite, so there will be a finite-sample bias.    

A natural question that arises is whether or not it is possible construct a bias-corrected estimate.   For nearest neighbor matching, \citet{abadie2011bias} constructed a bias-corrected estimator of the average causal effect.   Their methodology consisted of adjusting for the fitted values for each subject under the potential outcome function.    That will not work for our situation since in effect, the potential outcome function will be constant for every subject within treatment group for a given stratum.    Thus, the fitted value adjustment of the type described in \citet{abadie2011bias} cannot be used here.   

We instead propose a novel bias-correction approach that is inspired by the simulation extrapolation approach from the measurement error modeling literature~\citep{carroll1996asymptotics}, which is commonly abbreviated as SIMEX.  In measurement error modeling, the SIMEX approach fits a sequence of parameter estimates with a given measurement error and extrapolates a parameter estimate based on assuming the measurement error variance approaches zero.    For our setting, we will use $J$, the total number of strata, in an analogous way to the measurement error variance. Here is the outline for the approach.   
\begin{enumerate}
    \item Create a grid of values of $J$.
    \item For each value of $J$ on the grid, compute the estimated average causal effect and the variance.
    \item Fit a linear regression of the estimates as a function of $J^{-1}$.
    \item Use as the proposed estimate the predicted value at $J^{-1}$.  If the previous step was fit using the \textsf{lm} function in R and and saved as an object \textsf{foo}, then one would use \textsf{predict(foo,invJ = 0)} to obtain the estimate.
\end{enumerate}
The idea behind the proposed bias correction approach is that we expect unbiased causal effect estimation to occur when $J$, the number of strata, approaches infinity, or equivalently when $J^{-1}$ approaches zero.   
For a given value of $J$, our theory from section 3 yields that the limiting distribution is asymptotically normal.  Thus, we are making the assumption that the interpolated value will also have an asymptotic normal distribution.  A heuristic argument is that since the causal effect estimator in step 5 is a linear combination of the estimates at each value of $J$, it should also be asymptotically normal.   Formal justification of this result is beyond the current scope of the paper and is currently under investigation.

%% file: sections/examples.tex
For all the examples, we assume that for Wald statistics, we can apply the methods from Sections 3.4 and 3.6 so that a normal distribution under the null hypothesis holds.   In addition, we employ a significance level of 0.05. 
\subsection{Right-heart catheterization study}
\label{sec:RHC}
In this example, we apply the proposed methodology to data from the Study to Understand Prognoses and Preferences for Outcomes and Risks of Treatments (SUPPORT).  This is a very commonly used dataset from papers on causal inference methods.   SUPPORT is a multicenter observational trial that followed patients in critical care for prospective outcomes.  One major causal effect of interest in the SUPPORT study is whether or not right heart catheterization (RHC) has an effect on death within 30 days.  
Further information about the study can be found in~\cite{connors1996effectiveness}  The dataset contains information on 5735 patients, 2184 of whom received RHC.    In the original paper by~\cite{connors1996effectiveness}, they performed propensity score matching and found an increased risk of 30 day mortality associated with RHC treatment (odds ratio, 1.24; 95\% confidence interval = 1.03-1.49).

There are 50 confounders in the dataset.  When we attempted to match on all confounders using the method in ~\citet{iacus2011multivariate}, we were unable to get the algorithm to run or to create matched strata with treatment and control observations.  Next, we used the K-means based quantization to construct strata.   A sequence of K-means results with varying values of $K$ were fit, combined with the bias correction method described in \S~\ref{sec:biasc}.  This yielded an average causal effect estimate of -0.041 with a standard error of 0.01.  Given that the outcome is binary (1 for survival past 30 days, 0 for death within 30 days), our method reveals that RHC is associated with a causal risk difference of decreasing 30-day survival by 0.04; this effect is significant $(p = 0.003)$. 

We also applied the random forests-based approach for creating strata, as it allows for more distributional robustness in the confounders.    Based on the computed proximity distance matrix, we constructed confounders using hierarchical clustering varying the location where the dendrogram is cut.    We again apply the extrapolation method to get a causal risk difference of -0.021 with a standard error of 0.014.  Based on the Z-statistic, the random forest approach gives less significant answers to the k-means analysis, with a p-value of $0.13$.  However, the directionality of the causal effect from random forests aligns with that of k-means.  We have included the R code for the analysis in the Appendix.  

\subsection{Cesarean observational study}
\label{sec:EFM}

The next example is to evaluate the effect of electronic fetal monitoring (EFM) on cesarean section (CS) rates using data from Beth Israel Hospital from 14484 women who delivered between January 1970 and December 1975.  The data were published in~\cite{neutra1980effect}, and they identified several confounding factors: nulliparity, arrest of labor progression, malpresentation and year of birth.  The confounders are all binary.   Given no confounders are continuous, this is a situation where we expect the k-means algorithm to not work as well.  

We begin by applying coarsened exact matching.   The algorithm yielded a set of 45 strata, with one stratum containing only treated observations.   This yielded a risk difference estimate of $8.62 \times 10^{-5}$, with an attendant standard error of $0.005$ (p-value = $0.98$).   Next, we applied the proposed k-means clustering estimator for the risk difference.  This yields a causal risk difference of $0.004$ with a standard error of $0.005$ (p-value = $0.44$) .  Finally, the random forests-based causal effect estimate is $5.27 \times 10^{-4}$ with a standard error of $0.005$ (p-value = $0.92$).   We note that the magnitudes of the k-means and random forests estimator are substantially larger in magnitude relative to the original CEM estimate of the average causal effect.  We note that while the former two are bias corrected, the CEM estimate is not.   We also ran a separate analysis with k-means clustering using $K=45$ to match the number of strata in CEM, followed by the two-sample difference in means to match CEM.  This yielded an estimate of $2.2 \times 10^{-4}$ with a standard error of $0.005$.  Reasons for the discrepancy between the CEM and the other approaches include the lack of bias correction for CEM as well as variation in the grouping of observations into strata. Thus, while we find variation in the parameter estimates across the three methods, the direction of the effect estimate is the same.  In addition, all three methods do not show evidence for statistical significance for $\alpha = 0.05$.  We have included the code and results for these results in the Appendix.

%% file: sections/discussion.tex
The generalized coarsened confounding method discussed in this article can provide consistent estimators for the average causal effect with an asymptotic justification of the number of strata $J \equiv J_n \rightarrow \infty$ as the sample size tends to infinity.    As has been noted by certain authors~\citep{black2020trouble}, there is a bias in applying coarsened exact matching for real analyses with finite-sample datasets. We demonstrated this formally in Section 3. In order to address the induced bias, we have introduced a novel bias correction process inspired by the SIMEX procedure from~\citet{carroll1996asymptotics}. In summary, our generalized coarsened confounding methodology and bias correction process provides a workflow for approximately consistent casual effect estimation.  Future work will seek to provide asymptotic justification for the bias correction methodology.   

However, there are some issues that still need to be addressed in real-data analyses. The first involves how to choose the right variables to represent the population in matching. The effectiveness of matching usually relies on the number of observations retained after matching. The number of variables selected to represent the cohort should be evaluated since the number of matched observations will decrease as the number of matching variables increases.  

The choice of binning strategy is also critical here. How to find the balance between the similarity of cohorts and the generalizability in population is worth exploring further. In our work, we suggest a consistency properity with the number of strata and sample size both tending to infinity.   It is not directly obvious what to do in finite samples.  As an alternative to the CEM approach to finding strata, we introduced K-means algorithm for confounder strata clustering, as well as the random forests algorithm.  However, these two algorithms do not immediately output a measure of balance. In some articles, an $L_1$ statistic was developed as an indicator for balance~\citep{iacus2009cem}. This statistic ranges from zero to one, where zero refers to perfect balance with equal proportion of treatment and control observations in each stratum, while one refers to the mutually exclusive situation in each stratum. Based on our bias correction approach, we can simulate the range of $L_1$ values from zero to one and evaluate it with the different number of strata.  We leave this to future work.  We could also use this approach with another statistic named Least Common Support (LCS) which indicates the percentage of strata based on different number of observations~\citep{iacus2009cem}.

A reinterpretation of coarsened exact matching, described in \S 3.2, is used in the paper.  It exploits an encoding paradigm in which confounders are converted into code vectors, which index observations from which causal effects can be computed.  Recently, a full encoding-decoding paradigm for causal inference was described in~\citet{liu2024encoding}.  This leads into deeper connections between causal inference, information theory and machine learning that we intend to explore in future work.  


%% file: sections/appendix.tex
\subsection*{R code for the RHC  data example in Section 4.1.}

\begin{verbatim}
# RHC datasets
library(tidyverse)
library(dplyr)
library(ATbounds)

rhc_raw <- RHC

rhc_cleaning <- rhc_raw %>% 
  select(RHC, 
         survival, 
         age, sex_Female, edu, race_black,
         race_other,income1,income2,income3,
         wt0, hrt1, meanbp1, resp1, temp1,
         card_Yes, gastr_Yes, hema_Yes, meta_Yes, neuro_Yes, ortho_Yes, renal_Yes, 
         resp_Yes, seps_Yes, trauma_Yes,
         amihx, ca_Yes, cardiohx, chfhx, chrpulhx, dementhx, 
         gibledhx, immunhx, liverhx, malighx, psychhx, 
         renalhx, transhx, aps1, das2d3pc, scoma1, 
         surv2md1, alb1, bili1, crea1, hema1, paco21, 
         pafi1, ph1, pot1, sod1, wblc1)

rhc_cleaning %>%
  miss_var_summary()

# first, let's do MatchIt
library(MatchIt)
m.out1 <- matchit(RHC~age+edu+sex_Female+race_black+
                  race_other+income1+income2+income3+
                  wt0+ hrt1+ meanbp1+ resp1+ temp1+
                  card_Yes+ gastr_Yes+ hema_Yes+ meta_Yes+ neuro_Yes+ ortho_Yes+ renal_Yes+ 
                  resp_Yes+ seps_Yes+ trauma_Yes+
                  amihx+ ca_Yes+ cardiohx+ chfhx+ chrpulhx+ dementhx+ 
                  gibledhx+ immunhx+ liverhx+ malighx+ psychhx+ 
                  renalhx+ transhx+ aps1+ das2d3pc+ scoma1+ 
                  surv2md1+ alb1+ bili1+ crea1+ hema1+ paco21+ 
                  pafi1+ ph1+ pot1+ sod1+ wblc1,data=rhc_cleaning,method="cem",
                  estimand="ATE")

# does not work!!!

ace.km <- NULL
var.ace.km <- NULL
for (K in c(2,5,7,10,20,50)) {
strata <- kmeans(rhc_cleaning[,c(-1,-2)],centers=K)
clus <- strata$cluster
rhc_cleaning$stratum <- clus

n.tx <-  rhc_cleaning %>%
  filter(RHC == 1) %>%
  group_by(stratum) %>%
  summarise(n = n())

n.cont <- rhc_cleaning %>%
  filter(RHC == 0) %>%
  group_by(stratum) %>%
  summarise(n = n())

tmpwt <- (n.tx[,2]+n.cont[,2])/dim(rhc_cleaning)[1]
tx.mean <- rhc_cleaning %>%
  filter(RHC == 1) %>%
  group_by(stratum) %>%
  summarise(mean(survival))

control.mean <- rhc_cleaning %>%
  filter(RHC == 0) %>%
  group_by(stratum) %>%
  summarise(mean(survival))
ace.km <- c(ace.km,sum(tmpwt*(tx.mean[,2]-control.mean[,2])))
# Now the variance

tx.var <- rhc_cleaning %>%
  filter(RHC == 1) %>%
  group_by(stratum) %>%
  summarise(var(survival))

control.var <- rhc_cleaning %>%
  filter(RHC == 0) %>%
  group_by(stratum) %>%
  summarise(var(survival))

var.ace.km <- c(var.ace.km,sum(tmpwt^2*(tx.var/n.tx + control.var/n.cont)[,2]))
cat(K,"\n")

}
ace.km <- ace.km
var.ace.km <- var.ace.km
Jinv <- 1/c(2,5,7,10,20,50)
tmp1 <- lm(ace.km~Jinv)
ace <- predict(tmp1,list(Jinv=0)).  # -0.041
tmp2 <- lm(var.ace.km~Jinv)
var.ace <- predict(tmp2,list(Jinv=0)) #0.0001863553

# Rf
rhc_cleaning <- rhc_cleaning[,-53]
library(randomForest)
rf1 <- randomForest(x=rhc_cleaning[,c(-1,-2)],y=NULL,
                    ntree = 1000, proximity = TRUE, oob.prox = TRUE)
hclust.rf <- hclust(as.dist(1-rf1$proximity), method = "ward.D2")

ace.rf <- NULL
var.ace.rf <- NULL
for (K in c(5,7,10,20,30)) {
rf.cluster = cutree(hclust.rf, k=K)

rhc_cleaning$stratum <- rf.cluster

n.tx <-  rhc_cleaning %>%
  filter(RHC == 1) %>%
  group_by(stratum) %>%
  summarise(n = n())

n.cont <- rhc_cleaning %>%
  filter(RHC == 0) %>%
  group_by(stratum) %>%
  summarise(n = n())

tmpwt <- (n.tx[,2]+n.cont[,2])/dim(rhc_cleaning)[1]
  
tx.mean <- rhc_cleaning %>%
    filter(RHC == 1) %>%
    group_by(stratum) %>%
    summarise(mean(survival))

control.mean <- rhc_cleaning %>%
  filter(RHC == 0) %>%
  group_by(stratum) %>%
  summarise(mean(survival))
ace.rf <- c(ace.rf,sum(tmpwt*(tx.mean[,2]-control.mean[,2])))
# Now the variance

tx.var <- rhc_cleaning %>%
  filter(RHC == 1) %>%
  group_by(stratum) %>%
  summarise(var(survival))

control.var <- rhc_cleaning %>%
  filter(RHC == 0) %>%
  group_by(stratum) %>%
  summarise(var(survival))

var.ace.rf <- c(var.ace.rf,sum(tmpwt^2*(tx.var/n.tx + control.var/n.cont)[,2],na.rm=T))
cat(K,"\n")
}
Jinv <- 1/c(5,7,10,20,30)
tmp1 <- lm(ace.rf~Jinv)
ace <- predict(tmp1,list(Jinv=0)) # -0.02168539
tmp2 <- lm(var.ace.rf~Jinv)
var.ace <- predict(tmp2,list(Jinv=0)) # 0.0002077953

\end{verbatim}

\subsection*{R code for the cesarean data example in Section 4.2.}
\begin{verbatim}

library(ATbounds)

library(MatchIt)
m.out1 <- matchit(monitor~arrest+breech+nullipar+year,data=EFM,method="cem",
                  estimand="ATE")

stratum <- m.out1$subclass
EFM$stratum.cem <- stratum
wt.stratum.cem <- apply(table(stratum,EFM$monitor),1,sum)/dim(EFM)[1]
cem.stratum.tx <- NULL
cem.stratum.var <- NULL
for (i in levels(EFM$stratum.cem)) {
  subg <- EFM$stratum.cem == i
  tmpy <- EFM$cesarean[subg == "TRUE" & !is.na(subg)]
  tmptx <- EFM$monitor[subg == "TRUE" & !is.na(subg)]
  tmpace <- mean(tmpy[tmptx == 1]) - mean(tmpy[tmptx == 0])
  tmpvarace <- var(tmpy[tmptx == 1])/sum(tmptx == 1) +
    var(tmpy[tmptx == 0])/sum(tmptx == 0)
  cem.stratum.tx <-c(cem.stratum.tx,tmpace)
  cem.stratum.var <- c(cem.stratum.var,tmpvarace)
}

ace.cem <- sum(wt.stratum.cem*cem.stratum.tx)
varace.cem <- sum(wt.stratum.cem^2*cem.stratum.var,na.rm=T)

# K-means

EFM <- EFM[,c(-7,-8)]
ace.km <- NULL
var.ace.km <- NULL
K <- 45
tmpkm <- kmeans(EFM[,c(-1,-2)],centers=K)
clus <- tmpkm$cluster
wt.stratum.km <- apply(table(clus,EFM$monitor),1,sum)/dim(EFM)[1]
km.stratum.tx <- NULL
km.stratum.var <- NULL
for (i in 1:K) {
  subg <- clus == i
  tmpy <- EFM$cesarean[subg == "TRUE" & !is.na(subg)]
  tmptx <- EFM$monitor[subg == "TRUE" & !is.na(subg)]
  tmpace <- mean(tmpy[tmptx == 1]) - mean(tmpy[tmptx == 0])
  tmpvarace <- var(tmpy[tmptx == 1])/sum(tmptx == 1) +
    var(tmpy[tmptx == 0])/sum(tmptx == 0)
  km.stratum.tx <-c(km.stratum.tx,tmpace)
  km.stratum.var <- c(km.stratum.var,tmpvarace)
}
ace.km <- sum(wt.stratum.km*km.stratum.tx)
varace.km <- sum(wt.stratum.km^2*km.stratum.var,na.rm=T)

# Estimates
# K = 2: ACE = 0.04347558, Var = 2.270514e-05
# K = 5: ACE = 0.04300296, Var = 2.362881e-05
# K = 10: ACE = 0.008925207, Var = 2.438034e-05
# K = 20: ACE = 0.00460203, Var = 2.440448e-05
# K = 45:  ACE = 0.0002226695 Var = 2.495849e-05
# km.stra.tum.tx <- ifelse(is.nan(km.stratum.tx),0,km.stratum.tx)

# Bias correction
Jinv <- 1/c(2,5,10,20,45)
y <- c(0.04347558,0.04300296,0.008925207,0.00460203,0.0002226695)
tmp.lm1 <- lm(y~Jinv)
ace <- predict(tmp.lm1,list(Jinv=0)) # 0.00387

# Now for the variance
y <- c(2.270514e-05,2.362881e-05,2.438034e-05,2.440448e-05,2.495849e-05)
tmp.lm2 <- lm(y~Jinv)
var.ace <- predict(tmp.lm2,list(Jinv=0)).  # 2.477363e-05

# Rf
library(randomForest)
rf1 <- randomForest(x=EFM[,c(-1,-2)],y=NULL,
                    ntree = 1000, proximity = TRUE, oob.prox = TRUE)
hclust.rf <- hclust(as.dist(1-rf1$proximity), method = "ward.D2")

ace.rf <- NULL
var.ace.rf <- NULL
K <- 40
rf.cluster = cutree(hclust.rf, k=K)
wt.stratum.rf <- apply(table(rf.cluster,EFM$monitor),1,sum)/dim(EFM)[1]
rf.stratum.tx <- NULL
rf.stratum.var <- NULL
for (i in 1:K) {
  subg <- rf.cluster == i
  tmpy <- EFM$cesarean[subg == "TRUE" & !is.na(subg)]
  tmptx <- EFM$monitor[subg == "TRUE" & !is.na(subg)]
  tmpace <- mean(tmpy[tmptx == 1]) - mean(tmpy[tmptx == 0])
  tmpvarace <- var(tmpy[tmptx == 1])/sum(tmptx == 1) +
    var(tmpy[tmptx == 0])/sum(tmptx == 0)
  rf.stratum.tx <-c(rf.stratum.tx,tmpace)
  rf.stratum.var <- c(rf.stratum.var,tmpvarace)
}

ace.rf <- sum(wt.stratum.rf*rf.stratum.tx)
varace.rf <- sum(wt.stratum.rf^2*rf.stratum.var,na.rm=T)

# Estimates
# K = 5: ACE = 0.0120924, Var = 1.969936e-05
# K = 7: ACE = 0.01096943, Var = 2.055659e-05
# K = 10: ACE = 0.008772231, Var = 2.281092e-05
# K = 20: ACE = 0.004587131, Var = 2.487004e-05
# K = 40: ACE = 0.0001869272, Var = 2.500779e-05
 
# Bias correction
Jinv <- 1/c(5,7,10,20,40)
y <- c(0.0120924,0.01096943,0.008772231,0.004587131,0.0001869272)
tmp.lm1 <- lm(y~Jinv)
ace <- predict(tmp.lm1,list(Jinv=0)) # 0.0005276872

# Now variance
y <- c(1.969936e-05,2.055659e-05,2.281092e-05,2.487004e-05,2.500779e-05)
tmp.lm2 <- lm(y~Jinv)
var.ace <- predict(tmp.lm2,list(Jinv=0)) # 2.608433e-05

\end{verbatim}
